%% file: NLTE_OSAnd_86.tex
\documentstyle[epsf,twocolumn]{mn}
\include{mac}

\def\phx{{\tt PHOENIX}}
\begin{document}
\title[OS And 86]
{NON-LTE MODEL ATMOSPHERE ANALYSIS OF THE EARLY ULTRAVIOLET
SPECTRA OF NOVA OS ANDROMEDAE 1986}
\author[Schwarz et al.]
{Greg J. Schwarz,$^1$
Peter H. Hauschildt,$^2$
S. Starrfield,$^1$
E. Baron,$^3$ \cr
France Allard,$^4$ 
Steven N. Shore,$^5$
and G. Sonneborn,$^6$ \\
$^1$Department of Physics and Astronomy,
Arizona State University, Tempe, AZ 85287-1504 \\
E-Mail: \tt schwarz@hydro.la.asu.edu, 
\tt sumner.starrfield@asu.edu \\
$^2$Dept. of Physics and Astronomy,
University of Georgia, Athens, GA 30602-2451 \\
E-Mail: \tt yeti@hal.physast.uga.edu \\
$^3$Dept. of Physics and Astronomy,
University of Oklahoma,
440 W. Brooks, Rm 131, Norman, OK 73019-0225\\
E-Mail: \tt baron@phyast.nhn.uoknor.edu \\
$^4$Department of Physics,
Wichita State University,
Wichita, KS 67260-0032 \\
E-Mail: \tt allard@eureka.physics.twsu.edu \\
$^5$Department of Physics and Astronomy,
Indiana University South Bend,
1700 Mishawaka Ave, South Bend, IN 46634-7111 \\
E-Mail: \tt sshore@paladin.iusb.edu \\
$^6$Laboratory for Astronomy and Solar Physics, Code 681,
Goddard Space Flight Center, Greenbelt, MD 20771 \\
E-Mail: \tt sonneborn@fornax.gsfc.nasa.gov}

\maketitle

\begin{abstract}

We have analyzed the early optically thick ultraviolet spectra
of Nova OS And 1986 using a grid of spherically symmetric, 
non-LTE, line-blanketed, expanding model atmospheres and
synthetic spectra with the following set of parameters: 
$5,000\le$ T$_{model}$ $\le 60,000$K,
solar abundances,
$\rho \propto r^{-3}$,
$\vmax = 2000\kms$,
$L=6 \times 10^{4}\Lsun$, and a 
statistical or microturbulent velocity of 50 $\kms$.
We used the
synthetic spectra to estimate the model parameters corresponding to the
observed {\it IUE} spectra. The fits to the observations were then iteratively
improved by changing the parameters of the model atmospheres, in
particular T$_{model}$ and the abundances, to arrive at the best fits
to the optically thick pseudo-continuum and the 
features found in the {\it IUE} spectra.  

The {\it IUE} spectra show two different optically thick subphases.
The earliest spectra, taken a few days after maximum
optical light, show a pseudo-continuum created by overlapping
absorption lines.  The later observations, taken approximately 3 weeks 
after maximum light, show the simultaneous presence of 
allowed, semi-forbidden, and forbidden lines
in the observed spectra.

Analysis of these phases indicate that OS And 86 had solar metallicities
except for Mg which showed evidence of being underabundant by as much as
a factor of 10.  We determine a distance of 5.1 kpc
to OS And 86 and derive a peak bolometric luminosity of $\sim$
5 $\times$ 10$^4$ L$_{\odot}$.  The computed nova parameters provide
insights into the physics of the early outburst and explain the spectra
seen by {\it IUE}.  Lastly, we find evidence in the later
observations for large non-LTE effects of Fe\,{\sc ii} which, when included, 
lead to much better agreement with the observations.

\end{abstract}


\section{Introduction}

Novae occur in binary systems where the secondary fills its Roche lobe
and accretes mass onto a white dwarf primary.  The accreted mass 
collects on the white dwarf until the temperature and pressure 
at the interface between the core and envelope become
so great that a thermonuclear runaway (TNR) occurs. The liberation of 
energy from the TNR produces the nova explosion which ejects mass from
the white dwarf.
If the nova ejects enough mass, the gas will be optically thick during
its early evolution. Model atmosphere
analyses of the optically thick nova spectra can then be used
to determine many physical parameters of the nova ejecta including the energy
distribution, the model temperature, the velocity structure,
elemental abundances, and density distributions
as a function of time.  These results provide strong 
constraints on hydrodynamic calculations of the initial nova explosion
and give insights into the physical phenomena underlying the outburst.

In this paper, we present the results of modeling of the 
early optically thick phase of OS And 86 (Nova Andromeda 1986)
with Non-LTE (NLTE) model atmospheres.
In \S 2, we review the basic parameters of OS And 86
and derive the reddening and distance.  We do this
because the determination of the reddening to OS And 86 is 
critical to accurately modeling the nova spectra while the 
distance determines the absolute properties, e.g. the luminosity.
\S 3 discusses the {\it IUE} observations and the changes observed in
the spectra during the early optically thick evolution. 
\S 4 reviews the PHOENIX stellar atmosphere code, including
the recent addition of the NLTE treatment of Fe\,{\sc ii} and 
O\,{\sc i}.  We show the results of the model atmosphere synthetic
spectra's fits to the {\it IUE} spectra in \S 5.  The computed parameters 
for the best comparison of the {\it IUE} data is shown as a function of time
to illustrate the physics of the outburst.  In \S 6
we compare synthetic spectra calculated with 
Fe\,{\sc ii} in Non-LTE to those of synthetic spectra with 
Fe\,{\sc ii} in LTE and show that the later optically thick spectra
can only be fit by synthetic spectra with Fe\,{\sc ii} in Non-LTE. 
Concluding remarks are presented in \S 7.

\section{Determination of the basic parameters of OS And 1986}

OS And 86 was discovered by Suzuki (IAUC 4281)
on 1986 December 5.  Kondo and Kosai (IAUC 4282)
reported a position of $\alpha =$ 23$^h$ 9$^m$ 47.$^s$72 and
$\delta =$ +47$^\circ$ 12$\arcmin$ 0.$\arcsec$8 (equinox 1950)
corresponding to a galactic
longitude of 106.$^\circ$05 and a galactic latitude of -12.$^\circ$12.
Observations in the ultraviolet with the
International Ultraviolet Explorer ({\it IUE}) began on 1986 December 9,
with excellent temporal coverage until the discovery of SN1987A in
February 1987 limited the availability of {\it IUE}.

Optical maximum occurred between December 7.5 UT (Kikuchi et al. 1988) 
and December 8.94 UT (IAUC 4282) with an
apparent visual magnitude of approximately 6.3.  The ultraviolet
maximum occurred between December 13.9 UT and December 16.9 UT with an
integrated flux (1175 {\AA} - 3200 {\AA}) 
$\geq$ 6.8 $\times$ 10$^{-9}$ erg cm$^{-2}$ s$^{-1}$.  
OS And 86 had an optical t$_3$ (the time
it takes the nova's light curve to decline 3 magnitudes
below maximum) of 20 $\pm$ 1 days (Kikuchi et al. 1988).  
This makes OS And 86 a ``fast'' nova according to the 
speed classification defined by Payne-Gaposchkin (1957). 
The t$_{3}^{uv}$ (the t$_3$ time in the ultraviolet), 
which is usually greater then 
the t$_3$ in the optical, was about 60 days (Austin et al. 1990).

Andrillat (IAUC 4289) reported broad, intense emission lines of 
H$\beta$ through H$_{15}$ and Fe\,{\sc ii} (multiplets 27, 28, 37, 38,
42, 43) in spectra taken on December 15 to December 21.
The H$\beta$ lines showed a blue shifted absorption component with
a mean radial velocity increasing from 1190 
km s$^{-1}$ on December 15 to 1270 km s$^{-1}$ on December 
21.  This is in agreement with values obtained by 
Changchun et al. (1988). Based on the strong optical Fe\,{\sc ii} lines,
OS And 86 was a standard Fe\,{\sc ii} (Williams 1992), or CO type nova, 
suggesting that the nova occurred on the surface of a white dwarf 
composed of carbon and oxygen.

Kikuchi et al. (1988) observed changes in the polarization properties
of OS And 86 that they attribute to the formation of dust
around 15 to 20 days after optical maximum.  They report a 3.5 
mag decrease from their observations starting 20 days after 
visual maximum.  The photometric system used by Kikuchi et al. is
transformable to the standard UBV system but only when the radiation is 
dominated by continuum light, i.e., at maximum light.  
A transformation to a standard system is impossible as the nova evolves
toward an emission line dominated spectrum because the passband is
narrower than the standard V filter and emission lines such as 
H$\alpha$ are excluded.  Therefore, we have compiled a light 
curve for OS And 86 
from the IAU circulars and AAVSO visual observation (Mattei 1995)
(see Figure 1).  This light curve
clearly shows a dip starting in early January 1987 and lasting 
until mid March of 1987 when the light curve resumed its 
exponential decline.  This dip in the light curve is a
characteristic of the formation of an optically thick dust shell.
This is contrary to 
Gehrz's (1988) assertion, based on infrared photometry, that 
OS And 86 is a case where no substantial
dust shell developed.  The strength of the dip in the V band 
is 1.5 $\pm$ 0.5 magnitudes.  The rapid appearance of a dust shell
is surprising when compared to other novae with optically 
thick dust shells reviewed by Gehrz (1988) and Shore et al. (1994).  
The dust shell in OS And 86 appeared in half the time as the one in
V1370 Aql, a very fast (t$_3$ = 10 days) nova and $1/3$ the 
time of the slow novae FH Ser and LW Ser (t$_3$ = 67 and
55 days respectively).

\subsection{Extinction and Reddening corrections}

The determination of the external reddening and interstellar 
extinction curve for OS And 86 is necessary for accurately 
modeling the spectra.  
Due to the poor response of the LWP camera on the {\it IUE} satellite 
between 2000 and 2300 {\AA}, the 
2175 {\AA} feature cannot be used to determine the reddening.  This
wavelength region was also noisy in later optically thin spectra.  
There are, however, other ways to determine
the amount of reddening of a nova and we summarize them here:

\begin{enumerate}
\renewcommand{\theenumi}{(\arabic{enumi})}

\item From a literature search for all galactic classical novae with UBV
photometry, Van den Bergh and Younger (1987) determined 
that the intrinsic color 
of a nova at t$_2$ (the time to fall 2 magnitudes) is 
(B-V)$_o$ = -0.02 $\pm$ 0.04 magnitudes.  Milani et al. (IAUC 4306) 
report a (B-V) of +0.25 at t$_2$ corresponding to an E(B-V) of 
+0.27 $\pm$ 0.04.

\item The intrinsic color index of classical galactic novae at 
maximum is given by Allen
(1973) to be (B-V)$^{max}_o \approx$  0.2.  Krisciunas et al. (IAUC 4282) 
report a value of (B-V)$^{max}$ = +0.41 on December 9.3, 1986 (UT)  
which implies an E(B-V) of 0.21 magnitudes.  

\item Miroshnichenko (1988) analyzed the UBV photometry of more than 20
novae and found that the time almost immediately after maximum 
light, when the color index remains approximately constant (called
the ``stabilization stage'') can be used to find E(B-V).  The
average for novae in Miroshnichenko's 
study is (B-V)$_o$ = -0.11 $\pm$ 0.02 during the stabilization stage.  
From IAU circulars and the 
photometry of Kikuchi et al., we estimate that this stage
started around Dec. 18th (UT) and continued for 10 days.  The average 
(B-V) during this time is 0.23 magnitudes, implying an E(B-V) = 0.34.

\item A different approach is to analyze line of sight extinction toward
stars or galaxies in the direction
of the nova.  Austin et al. (private communication) 
used the color excess map compiled by Burstein and Heiles (1982) to show that 
OS And 86 lies in a region where E(B-V) = 0.24 magnitudes.
The color excess map is a combination of galactic and 
extragalactic reddening and {\it does not} include 
any circumstellar extinction .

\end{enumerate}

The mean reddening of the four values is 0.26 magnitudes.  
For OS And 86, therefore, we adopt an estimate of E(B-V) =
0.25 $\pm$ 0.05.

\subsection{Absolute Magnitude and Distance Determination}

A determination of the distance is essential for determining absolute
properties of a nova.  A great deal of effort has been expended on 
determining M$_V$, and thus the distance, of novae 
from samples of novae with independently known distances.  
A compilation of maximum magnitude versus the rate of decline (MMRD)
relationships is published in Chochol et al. (1993) for V1974 Cyg 1992.
The relations are: 
\begin{enumerate}
\renewcommand{\theenumi}{(\arabic{enumi})}
\item M$_{0,V}$ = -10.70 + 2.41 log(t$_2$) (Cohen 1985),
\item M$_{0,V}$ = -7.89 - 0.81 arctan((1.32 - log(t$_2$))/0.19) (Capaccioli et al. 1989), 
\item  M$_{0,V}$ = -11.75 + 2.5 log(t$_3$) (Schmidt 1957), 
\item M$_{15,V}$ = -5.6 $\pm$ 0.14 (Cohen 1985), 
\item M$_{15,V}$ = -5.23 $\pm$ 0.39 (van den Bergh and Younger 1987),
\end{enumerate}
\noindent where M$_{0,V}$ and M$_{15,V}$ are the absolute V magnitudes 
at maximum and 15 days after maximum, respectively.
OS And 86 had a t$_2$ of 8 $\pm$ 1 days.  The mean absolute
magnitude at maximum light from the first three methods is M$_{0,V}$ = -8.6.  
If we assume an
extinction of A$_v$ = 3.1E(B-V) = 0.78, then the distance to OS And is 
6.6 kpc.  A visual inspection of the lightcurve shows
that OS And 86 had V $\approx$ 9 mag fifteen days after optical maximum.  
Using the same extinction as before and the mean M$_{15,V}$ from the 
last two methods, we find a distance to OS And 86 of 5.3 kpc.

The MMRD relationships depend on a statistical fit to surveys 
containing different composition classes and different speed classes
of novae.
These methods work well in determining an {\it average} relationship 
for all novae in the sample but when applied to an {\it individual}
nova they cannot be very accurate because of the large 
spread in the properties of individual novae. 

There is another approach to determine the distance to novae which
is {\it not} based on statistical methods (Starrfield et al. 1992a). 
We require a similar nova at a known distance 
and with a known extinction and reddening.
Novae in the LMC are a good choice because
7 different novae have been observed in the LMC with {\it IUE}, it is
at a well established distance, and the extinction 
and reddening to the LMC is small.  One fast CO type LMC nova, 
LMC 1992 (hereafter LMC 92), exhibited an outburst 
very similar to OS And 86, with the exception that LMC
92 was faster with t$_3$ = 16 $\pm$ 2 days.  
We have obtained an {\it IUE} composite spectrum of LMC 92 on Nov. 20, 1992 
(SWP46299+LWP24328) at 
approximately the same stage of spectral development as OS And 86 on 
Dec. 13, 1986, (see Figure 2).  The LMC 92 spectrum displays most of
the same features and mimics the shape of the pseudo continuum of
OS And 86.  If we assume that the UV luminosity
of OS And 86 was the same as LMC 92 at this epoch,
then the distance to OS And 86 is given by:
\[
D{_{OS}}=D{_{92}} \sqrt{ \frac{f_{92}}{f_{OS}}},
\]
where $f_{92}$ and $f_{OS}$ are the total dereddened UV fluxes 
(1175 {\AA} to 3300 {\AA})
of LMC 92 and OS And 86 respectively, (dereddened with an 
E(B-V) of 0.15 $\pm$ 0.05
and 0.25 $\pm$ 0.05).  Gould (1995) reports a value of 
47.3 $\pm$ 0.8 kpc to the centre of the LMC.  We allow an
additional $\pm$ 500 pc due to the uncertainty in the location of 
LMC 92 within the LMC and adopt D$_{92}$ = 47.3 $\pm$ 1.3 kpc.
The total observed UV flux of LMC 92 and OS And 86 are 
3.9 $\pm$ 1.7 $\times$ 10$^{-10}$ and
3.3 $\pm$ 1.3 $\times$ 10$^{-8}$ ergs s$^{-1}$
cm$^{-2}$, respectively.  Using the flux ratio we determine a 
distance of 5.1 $\pm$ 1.5 kpc.  Here the dominant source of error is the
uncertainties in the reddening.  This distance places OS And 86 at a
distance of 1.1 $\pm$ 0.3 kpc below the galactic plane.  This is 
outside the galactic disk and suggests that OS And 86 is not from the 
young disk population.

Using the distance determined from this technique
we find that OS And 86 had M$_{0,V}$ = -8.0 $\pm$ 0.7.
The mass of the underlying white dwarf can then be estimated from the formula
derived by Livio (1992):
\[
{\rm M}_{WD} \approx 10^{(\frac{-8.3 - {\rm M}_{0,B}}{10.0})}, 
\]
where M$_{0,B}$ is the absolute B magnitude at visual maximum  
and M$_{WD}$ is the mass of the white dwarf in solar masses.  
The equation is an approximation since it is only a 
function of the absolute B magnitude and it neglects
effects such as the white dwarf's luminosity, magnetic field, the mass
accretion rate, and the composition of the accreted material.  
Nevertheless, Livio's equation gives an estimate of the white dwarf mass
of 0.9 $\pm$ 0.2 M$_{\odot}$.  
Table 1 summarizes the basic parameters of OS And 86.

\section{Observations}

We have retrieved high and low resolution archival {\it IUE}
spectra of OS And 86 obtained with the  
long wavelength primary (LWP: 2000-3300 {\AA}) and the short wavelength
primary (SWP: 1175-1950 {\AA}) cameras.  These spectra were
reduced at Goddard Space Flight Center (GSFC) Regional Data Analysis 
Facility (RDAF) using the
standard {\it IUE} software and special purpose IDL routines.  

{\it IUE} took 3 high (R = 10$^4$) dispersion spectra and 29 low
(R $\approx$ 300) dispersion spectra of OS And 86 during the first month
after discovery while the nova atmosphere was still optically thick.
Unfortunately, not all of these spectra are suitable for analysis.
Due to the {\it IUE} satellite's small dynamic range, some spectra have
their strongest features overexposed,
while some shorter exposure spectra underexpose the weakest regions. 
On a few days, two low resolution LWP and SWP 
spectra were taken within one hour of each other.  One of each
low resolution pair was a long 
exposure while the other two were short exposures.  This was done
to compensate for the limited dynamic range of {\it IUE}.  Therefore,
we combined the best exposed portions of each spectrum to
provide a final spectrum for that particular time.  Table 2 
gives the dates on which suitable low resolution 
{\it IUE} spectra exist and the image numbers and exposure information
of each spectrum.

The optically thick phase of a nova outburst can be divided into three
distinct phases.  These are the ``fireball'' phase (Gehrz 1988, Shore et
al. 1993), the ``continuous mass loss'' phase 
(Hauschildt et al. 1994), and the ``pre-nebular'' phase.  

The ``fireball'' is the first stage in a nova's development 
and marks the phase
when the ejected material is adiabatically expanding and cooling
from very high initial temperatures.  The cooling of this optically
thick material shifts the flux peak from the UV to optical wavelengths
and causes a steep decline in the ultraviolet.  Because the fireball ejecta
become optically 
thin before maximum light in the V band, it is difficult to 
obtain spectra during this relatively rapid phase of development of
the nova.  Only for novae discovered very early during the
rise to maximum in V, such as V1974 Cyg 
(Hauschildt et al. 1994; Shore et al. 1993) and Nova LMC 1991 
(Schwarz et al. in preparation), have this interesting ``fireball'' phase been
observed.  OS And 86 was probably just past the fireball phase at 
the time of the first {\it IUE} observations.  

The fireball ejecta become progressively more transparent and the deeper
layers of the nova atmosphere become visible as the optical flux
passes through maximum.  Previous work (Hauschildt et al. 1994, 1995a) on
Nova Cas 1993 has  
shown that nova atmospheres have sufficient radiation pressure 
to drive mass loss at this epoch while the optical emission lines show
strong P-Cygni profiles, indicative of a continuous outflow.  We, 
therefore, term this the ``continuous mass loss'' phase.
It is around this time that the ``iron curtain'' comes to dominate
the spectrum. The temperature drop
from the ``fireball'' phase causes higher ions
to recombine, which in turn causes a large increase in UV opacity.
This explains why the ``continuous mass loss'' phase is characterized by
a pseudo-continuum created by overlapping absorption lines, mainly 
Fe\,{\sc ii}.  The pseudo-continuum peaks strongly at longer wavelengths 
($\approx$ 3000 {\AA}) with almost undetectable flux at the 
shorter wavelengths where the opacity is largest.  

The first two spectra in Figure 3 and Table 2 show OS And 86 
in this early ``continuous mass loss'' phase.  
These spectra are very similar in 
appearance except that the integrated flux is about 30\% higher on 
Dec. 11.  There is almost no flux shortward of 1500 {\AA} on either
day.  The spectra are dominated by features that appear to be emission
lines, but these ``lines'' are in reality {\it regions of transparency 
where the opacity is reduced.}  The features between 2600 {\AA} and 
2700 {\AA}, and between 2900 {\AA} and 3000 {\AA},
are gaps in the Fe\,{\sc ii} line absorption (Hauschildt 
et al. 1994).  Mg\,{\sc ii} 2800 {\AA} (and possibly Al~III 1860 {\AA})
is the only true emission line which is present
during this epoch and it is strongly blended with Fe\,{\sc ii} lines.

As this phase progresses, the overall slope in the 
ultraviolet spectra evolves so that the pseudo-continuum 
gradually appears to flatten.  The drop in density from the expansion
causes a decrease in line 
opacity in this spectral region.  This ``lifting of 
the iron curtain'' (Hauschildt et al, 1992b, 1994, Shore et al. 1994) 
causes the flux peak to gradually shift into the SWP region of 
the spectrum (see \S 5). 

The ``lifting of the iron curtain'' in OS And 86 is shown in the middle 
two spectra in Figure 3 and in Table 2.  These spectra show an increase 
of about a factor of 3 in integrated flux below 2000 {\AA} 
with respect to the first 
two spectra.  The increase is so great that the spectrum above 1700
{\AA} is overexposed on Dec. 16th.  The dramatic increase in the 
SWP region and the slight increase in total integrated 
flux of the LWP give the impression 
that the spectra have flattened.  Table 2 shows that at this time
OS And 86 reached its maximum in the UV.  This is directly attributable to
the increase in radius of the ejected gas and subsequent 
drop in Fe\,{\sc ii} opacity.

The ``pre-nebular'' phase is characterized by the emergence of
moderately ionized ($\la$ 54 eV ionization potential from the 
presence of He\,{\sc ii}), allowed, 
semi-forbidden, and forbidden emission lines superimposed on
a pseudo continuum that is peaked toward the blue. 
These lines are formed in the outer region of the atmosphere where the
density is sufficiently low for nebular emission lines to appear.
As the opacity continues to drop, these lines 
strengthen and the spectra begin to resemble those obtained during the
optically thin ``nebular'' phase of other novae.  

The last 2 spectra in Figure 3 and Table 2 show this phase in OS And 86.  
Their psuedocontinua are strongly sloped to the blue
and are dominated by pre-nebular lines.  The strongest of the lines are
due to O\,{\sc i}, N\,{\sc i}, He\,{\sc ii}, C\,{\sc ii}, 
N\,{\sc ii}, Mg\,{\sc ii}, and N\,{\sc iii}.  
A complete list of the pre-nebular lines is given in Table 3.  
At this phase, the strongest lines are from CNO ions with 
low ionization states such as O\,{\sc i} (1304 {\AA}), C\,{\sc ii}
(1335 {\AA}), and N\,{\sc iii} (1750 {\AA}).  In addition, there 
is evidence of emission lines to the metastable
$^2$P$^0$ and $^2$D$^0$ states of N\,{\sc i} during this
phase.  The strongest emission lines from N\,{\sc i} (Moore 1993) are:

\begin{enumerate}
\renewcommand{\theenumi}{(\arabic{enumi})}

\item 3d($^2$D) - 2p$^3$($^2$P$^0$) at 1310 {\AA},
\item 3d($^2$P) - 2p$^3$($^2$P$^0$) at 1319 {\AA},
\item 3s($^2$D) - 2p$^3$($^2$P$^0$) at 1411 {\AA},
\item 3s($^2$P) - 2p$^3$($^2$P$^0$) at 1743 {\AA},
\item 3s($^2$D) - 2p$^3$($^2$D$^0$) at 1243 {\AA} and,
\item 3s($^2$P) - 2p$^3$($^2$D$^0$) at 1493 {\AA}.

\end{enumerate}

Figure 4 shows the N\,{\sc i} emission lines in the Dec. 27th high 
resolution UV spectrum, SWP29981.  
The line at 1319 {\AA} is clearly seen while the line at 1310 {\AA}
is blended with O\,{\sc i} 1304 {\AA} (Figure 4a).  Interstellar 
absorption lines of O\,{\sc i} and C\,{\sc ii} 1336 {\AA} can also be seen.
In Figure 4b, the line at 1411 {\AA} is blended with an unidentified
line at 1415 {\AA}.  The O\,{\sc iv}] and Si\,{\sc iv}] lines
around 1401 {\AA} are present but weak and 
severely blended.  The location of the line centre in Figure 4c indicates  
that the strong line near 1486 {\AA} is {\it not} 
N\,{\sc iv} but rather N\,{\sc i} at 1493 {\AA} (see Scott et al.
1995).  The C\,{\sc iv} line at 1550 {\AA} is present but is also weak.
In Figure 4d, the N\,{\sc iii}] line at 1750 {\AA} is clearly
blended with the N\,{\sc i} line at 1743 {\AA}.

\section{Model Atmospheres}

\subsection{Model construction}

The spectral syntheses of the early spectra of OS And 86
were calculated using the method described by
Hauschildt and Baron (1995b). Therefore, here we provide 
only a brief description of the method and summarize 
the recent changes to  Hauschildt's non-LTE, 
expanding, stellar atmosphere code \phx.

\phx\ solves the special relativistic equation of radiative 
transfer (SSRTE) in the Lagrangian frame self-consistently with 
the multi-level, non-LTE rate equations and the special relativistic 
radiative equilibrium (RE) equation in the Lagrangian frame.
Numerical methods used in \phx\ include the following: 
(i) the solution of the SSRTE is done using 
the operator splitting method described by Hauschildt (1992a),
(ii) the RE equation is solved by an Uns\"old-Lucy type temperature
correction scheme (Allard 1990), and (iii) the multi-level non-LTE
continuum and line transfer problem is treated using the 
operator splitting method described by Hauschildt (1993). 
 
 The following species are treated in non-LTE: H~I (10 levels),
Mg\,{\sc ii} (18 levels), Ca~II (5 levels), Ne~I (26 levels), 
and O\,{\sc i} (36 levels) (Hauschildt \etal\ 1994).
The lines are represented by depth-dependent Gaussian
profiles with 25 wavelength points per permitted non-LTE line. 

A recent addition to the \phx\ atmosphere code 
is a fully non-LTE treament of Fe\,{\sc ii}.  This ion  
plays an important role in the formation of early nova spectra
because of its high abundance and low ionization threshold.  We used an 
Fe\,{\sc ii} model atom that includes 617 levels, over 10$^4$ primary 
permitted transitions, and over 10$^6$ secondary transitions
(Hauschildt and Baron 1995b; Hauschildt et al. 1996).  
Its inclusion considerably alters our synthetic 
spectra and dramatically improves the fits to the {\it IUE} 
data when compared to LTE 
treatments.  This will be demonstrated in \S 6. 

In addition to the non-LTE lines, the models self-consistently include 
line blanketing by the most important ($\approx 10^6$) 
metal lines selected from the Kurucz (1994) line list.  
The entire list contains close to 42 million
lines; however, not all of them are important for a particular nova model.
Therefore, before each temperature iteration, a subset
is selected from the original list by a 
process described in Hauschildt et al. (1992b, 1994). 
We treat line scattering in the metal lines of LTE species
(approximately) by parameterizing the albedo for single scattering, 
$\alpha$. The detailed calculation of $\alpha$ would require a full
non-LTE treatment for {\sl all} lines
and continua, which is outside of the scope of this paper.  
Tests have shown that the line profiles do
not depend sensitively on $\alpha$ as a direct result of the
velocity gradient in nova photospheres and that our approach is 
reasonable. An
average value of $\alpha=0.95$ is used; recent results indicate that 
this is an acceptable choice (Hauschildt and Baron 1995b).
The continuous absorption and scattering coefficients are
calculated using the species and cross sections 
described in Hauschildt et al. (1995a) and Allard and Hauschildt
(1995).  

\subsection{The model parameters}

The model atmospheres are characterized by the following 
parameters (see Hauschildt et al. 1992b for details): 
\begin{enumerate}
\renewcommand{\theenumi}{(\arabic{enumi})}
\item the reference radius $R$, which is the radius where either
the continuum optical depth in absorption or extinction at 
5000$\ang$ is unity,
\item the model temperature T$_{model}$\unskip, which is defined by
means of the luminosity, $L$, and the reference radius,
$R$, (T$_{model}=(L / 4\pi R^2 \sigma )^{1/4}$ where $\sigma$ is
Stefan's constant),
\item the density parameter, $N$, ($\rho(r)\propto r^{-N}$),
\item the maximum expansion velocity
given by $v = \dot M /$ 4$\pi r^2 \rho$ with the mass loss
rate, $\dot M$(r), assumed to be a constant, 
\item the density, $\rho_{\rm out}$, at the outer edge of the envelope, 
\item the statistical velocity $\xi$, treated
as depth-independent isotropic turbulence, and
\item the element abundances.  
\end{enumerate}

We emphasize that for {\it extended} model atmospheres, one should not 
assign a  physical interpretation to
the parameteric combination of T$_{model}$ and $R$.  Previous work 
(Hauschildt et al. 1992b, 1994, 1995a) has called the model temperature an
``effective temperature'', or T$_{eff}$, but this is technically not
correct.  In plane-parallel stellar atmospheres, it is possible to define an
effective temperature as the temperature of a black body emitting the
equivalent flux.  However, in an extended atmosphere there is no longer
a {\it unique radius} at which this can be defined.  By
using a reference radius at a prescribed {\sl continuum}
optical depth scale at $\lambda=5000\ang$  we define a model
temperature.  We emphasize that 
{\it the model temperature must be regarded only as a convenient
numerical parameter used to describe the model 
and is not directly comparable to any observationally
determined radius except at 5000 {\AA}.}  Pistinner et al.
(1995) present a detailed discussions of nova atmosphere parameterization.

\section{Results of NLTE modeling}

The parameters that affect the synthetic spectra most sensitively are
the model temperature, the density parameter, and the metal (Z $>$ 2)
abundances.  Previous work (Hauschildt et al. 1992b: Hauschildt et al. 1994)
and hydrodynamic calculations of nova outbursts (Starrfield et al. 1992b) have
shown that the post optical maximum optically thick phases are best modeled
with N $\approx$ 3.  With N fixed at 3, we created three libraries of
synthetic spectra where only the model temperature was varied.
Each library consisted of models with metal abundances (by number) 
of 0.5, 1, and 2 times the solar value. 
We did this for two reasons.  The metal rich synthetic spectra 
were calculated because nova theory predicts that metals (namely CNO)
should be enhanced relative to hydrogen because of mixing 
of accreted material with core white dwarf material (Politano et al.
1995).  To investigate the possibility
that the secondary star of OS And is a metal poor subdwarf since it
lies outside the galactic disk, we created the metal poor synthetic spectra.

A maximum velocity of $v_{max}$ = 2000 km s$^{-1}$ was chosen as a
reasonable guess for a typical nova, while a statistical velocity of
$\xi$ = 50 km s$^{-1}$ was chosen as a typical value for hot stars.
The model's luminosity was chosen as 6 $\times$ 10$^4$ L$_{\odot}$ 
(see section \S 2.2).
The outer pressure, $\rho_{\rm out}$, was set to 10$^{-3}$ dyn cm$^{-2}$
to ensure that the material above the model atmosphere was optically
thin at all wavelengths.  The synthetic spectra were convolved 
with a gaussian kernel with 5 {\AA} resolution to simulate the low
resolution {\it IUE} spectra. 

In Figure 5, we show a collection of synthetic spectra compared to 
the {\it IUE} spectrum of Dec. 11th.  Figures 5a,b, and c show the best 
fitting synthetic spectrum with the metal rich, solar, and metal 
poor abundances, respectively.  Notice that increasing the 
metallicity (Figure 5a) produces a spectrum that has a higher 
model temperature (19000 K) than the solar abundance synthetic 
spectrum (17000 K: Figure 5b) which in turn is hotter than the metal poor 
synthetic spectrum (16000 K: Figure 5c).  We will explain this
phenomenon shortly and describe how it can be used to determine the model
temperature of the nova.  

Although these three synthetic spectra fit the {\it IUE} spectrum 
well at most wavelengths, below 1500 {\AA} they  
predict about 100 times the observed flux.
In principle, this region of the spectrum can be used to determine
the CNO abundances because of the large opacity from the numerous CNO
lines located below 1500 {\AA}.  Increasing the CNO abundance reduces
the flux in the synthetic spectra in this
spectral region.  However, our LTE treatment of CNO does not 
significantly improve the fits of the synthetic spectra with CNO abundances 
greater than 10 times solar.  An accurate analysis of
this spectral region requires that all CNO ions be treated in NLTE which
is currently being implemented.  Therefore, we {\it cannot} 
presently say by how much the CNO elements may be overabundant.  

Most of the spectral features are 
reproduced by the synthetic spectra in the region between 1500 
and 2600 {\AA}.  The flux in the synthetic spectra 
is generally too low by as much as 40$\%$ between 
2300 and 2600 {\AA}.  We {\it stress} that in this region, it is only 
important for the synthetic spectra to show the same features as seen in
the {\it IUE} spectrum and {\it not} to precisely reproduce the flux.
This is because the exact shape of the interstellar extinction
curve, particularly the strength of the 2175 {\AA} absorption feature,
is not known for OS And 86.  

The features above 2600 {\AA}, caused by the gaps in the distribution 
of iron peak absorption lines, are well fit by all of the 
synthetic spectra.  All three synthetic
spectra predict too much flux at Mg\,{\sc ii} (2800 {\AA}).   
A synthetic spectrum with the magnesium abundance 
(relative to solar) {\it reduced by a factor of 10} was 
found to substantially improve the fit to the Mg\,{\sc ii} emission line.
Because of the strength of this resonance transition we cannot 
determine the magnesium abundance with high precision, but it seems
likely that magnesium is depleted relative to hydrogen; possibly
by as much as a factor of 10 (from a solar abundance).

The reason that all three synthetic spectra exhibit a good 
comparision to the {\it IUE} data is the relative insensitivity
of the iron curtain to the details of the model.  In Figure 6a, we plot the
metal rich synthetic spectrum with a model temperature of
17000 K (dotted line) and a solar metallicity synthetic spectrum with 
the same model temperature (solid line).  The higher metallicity spectrum
shows stronger metal lines, mostly Fe, in the optical as compared to
the solar metallicity spectrum.  However, in the UV, the 
presence of the iron curtain implies that a synthetic spectrum with 
larger metal abundances has an increased opacity. This produces a 
steeper UV pseudo continuum, which can be seen in the flux ratios.  
In order, to produce a UV synthetic spectrum which 
resembles the solar metallicity spectrum at 17000 K (dotted line) 
but using a higher metallicity, we must increase the 
model temperature by 2000 K (solid line: see Figure 6b).  
The depopulation of Fe\,{\sc ii} by the hotter radiation field is 
balanced by the increase in Fe abundance.  
Even though different combinations of model temperature and metallicity
produce UV spectra that are qualitatively similar, Figure 6b shows 
clearly that the
optical continuum {\it relative} to the UV is very different between the 
two spectra.  Unfortunately, there are no flux calibrated optical
spectra for OS And 86 during the optically thick epoch to help us 
determine the metallicity.  This underscores the need to have flux
calibrated {\it optical} observations together with the UV for novae.

We cannot presently say which of the three synthetic spectra is most
representative of OS And 86 (from this {\it IUE} spectrum alone), 
since the comparisons
are essentially the same in these low resolution {\it IUE} spectra.  
To resolve the ambiguity, we used high resolution {\it IUE} spectra.
In Figure 7, we 
show the {\it IUE} high resolution spectrum (LWP9719 (2400 to 2780 {\AA}) +
LWP9717 (2780 to 3200 {\AA})) of December 16th, 1986 as a dashed line to
facilitate viewing.
Figures 7a,b, and c show the best fitting synthetic spectra with 
2 (T$_{model}$=20000K), 1 (T$_{model}$=20000K), and 0.5 (T$_{model}$=19000) 
times solar metallicities respectively (solid line).
The twice solar synthetic spectrum produces absorption features between 
2850-2900 {\AA} and between 2950-3250 {\AA} that are stronger than 
observed.  The Fe\,{\sc ii} absorption in the wing of Mg\,{\sc ii}, at 
2750 {\AA}, is too weak, the 
flux predicted blueward of 2600 {\AA} is too low, and the flux longer
then 3250 {\AA} is too high, compared to the {\it IUE} spectrum.  These results
force us to abandon the high metallicity models for OS And 86.

The other two synthetic spectra fit most of the features fairly 
well and reproduce the flux throughout the {\it IUE} spectrum
except for the Mg\,{\sc ii} 2800 {\AA} line. We suspect that at this
epoch the Mg\,{\sc ii} line includes an additional 
component from the outermost optically thin regions of the atmosphere.
This component is not currently included in the model calculations and
an analysis of Mg\,{\sc ii} at this epoch would have to account for this
``pre-nebular'' contribution.
A careful examination of Figures 7a and b shows that the
synthetic spectrum with solar metallicities gives a slightly better fit.
The features at 2550 {\AA}, 2610 {\AA}, 2675 {\AA}, and the flux
redward of 3000 {\AA} are in better agreement with the observed {\it IUE} 
spectrum.  Although we adopt the synthetic spectra library with solar
metallicity for the rest of the discussion,
we point out that the metal poor model is not inconsistant with 
the {\it IUE} data alone.

\subsection{Time development of the model atmosphere}

Each model atmosphere calculates nova properties at 50
logarithmically spaced depth points between $\tau_{std}$ = 10$^{-6}$
and $\tau_{std}$ = 10$^4$, where $\tau_{std}$ is the optical depth of 
the continuum at 5000 {\AA}. 
We use the information contained in the $\tau$ grids, from the 
best fit model atmospheres, to illustrate 
the physics in optically thick UV nova spectra.

The optical depth, electron temperature and density 
are presented at a fixed radius of 10$^{13}$ cm as a function of
time in Figures 8a, b, and c.  The figures show that
as the nova evolves, the optical depth decreases, the 
electron temperature increases, and the density drops at this radius.
The interpretation is that as the nova expands, the
density drops and the deeper, hotter layers are exposed.  The outer
layers decrease in density due to expansion and
are exposed to a hotter radiation field allowing the strong
``pre nebular'' emission lines to appear superimposed on top of the
pseudo continuum.  

The ``lifting'' of the iron curtain is illustrated in Figure 8d.  This
Figure shows the Fe\,{\sc ii} number density as a function of time at
$\tau_{std}$ =  1 in the solar abundance model atmosphere.
During the earliest epochs of the ``continuous mass loss''
phase, the number density of Fe\,{\sc ii} is high, which produces the
increased opacity in the UV spectrum.
The Fe\,{\sc ii} number density falls rapidly during expansion,
which is manifested in the observed UV spectra as a flux increase
in UV.

In Figure 8e, we show the bolometric flux.  
It has been shown (Hauschildt et al. 1995a) that the 
synthetic spectra are insensitive to the luminosity of the model 
atmosphere and thus we cannot determine the luminosity from the
synthetic spectra alone.  We can find the
bolometric flux of OS And through another method.  First, we note that 
the synthetic spectrum, which best fits each {\it IUE} spectrum, 
includes regions outside the wavelength regime of {\it IUE}.  
By summing the flux in each synthetic spectrum, 
we arrive at a bolometric flux for each epoch of the {\it IUE} data.  
The plot
shows that the flux was constant, within the limits of our error,
for about the first week after visual maximum when it then declined.
This decline is due to the presence of strong pre-nebular
emission lines in the UV and the optical (Changchun et al
1987) beginning in late December.  Since these lines are not included
in the synthetic spectrum's bolometric flux, and are a 
significant contribution to the flux (of order 10$\%$ in the UV alone), 
the last two data points are only
lower limits.  Using the distance determined in \S 2, the 
bolometric luminosity of OS And 86 is 5 $\pm$ 2 $\times$ 10$^4$ 
L$_{\odot}$ or about the Eddington limit for a 1 M$_{\odot}$ 
white dwarf.  This is in agreement with the white
dwarf mass derived in \S 2.

In addition, the constant bolometric flux of the early optically
thick nova is consistant with the reddening determined in \S 2.
If the luminosity is constant, then the maximum integrated flux 
in the optical is equal to the maximum integrated 
flux in the UV.  We use photoelectric B and V magnitudes, converted 
to fluxes using Allen (1973), on 1986 December 9.34 (IAUC 4282)
to approximate the maximum integrated optical flux. 
The maximum integrated UV flux is equal
to the maximum optical flux when E(B-V) $\approx$ 0.25.

\section{Fe\,{\sc ii} NLTE vs Fe\,{\sc ii} LTE} 

In LTE models, the occupation numbers, the opacity,
and the emissivity are assumed to be locally in thermodynamic equilibrium 
throughout the atmosphere.  Generally, while the assumption of LTE
should be acceptable in stars, an accurate treatment for nova 
atmospheres demands that the most important species be treated in NLTE.
This is because nova atmospheres have large temperature and density gradients, 
low densities, and highly non-Planckian radiation fields and 
thus {\it must} exhibit non-LTE behavior. 

Fe\,{\sc ii} LTE model atmospheres with model temperatures 
$\approx$ 17,000 K produce synthetic spectra that are very similar
to their NLTE model counterparts.  At this model temperature,
the departures from LTE are not significant (see Hauschildt et al. 1996).
However, in hotter models, the NLTE effects in Fe\,{\sc ii} are 
considerable.  The hotter radiation field prevents the recombination
of Fe\,{\sc iii} to Fe\,{\sc ii}, thus decreasing the Fe\,{\sc ii} 
abundance.  Figure 9a and b give the LTE (triangles) versus the 
NLTE Fe\,{\sc ii} (diamonds) number density as a function 
of optical depth for the best fit synthetic spectra on Dec. 11th 
(T$_{model}^{LTE}$ = T$_{model}^{NLTE}$ = 17000 K) and Dec. 27th 
(T$_{model}^{LTE}$ = 27000 K and T$_{model}^{NLTE}$ = 25000 K).  
The cooler model atmospheres show essentially the same Fe\,{\sc ii} 
abundances in the outer atmosphere regardless of the 
way Fe\,{\sc ii} is treated. 
In the hotter models, however, the LTE model shows a severe overabundance of
Fe\,{\sc ii}, by as much as 10$^3$, in the outer regions of the atmosphere.  
The consequences of the increased Fe\,{\sc ii} number density can be seen 
in the synthetic spectra.

In Figure 10a and b, we show the Dec. 27th {\it IUE} spectrum 
compared to the solar abundance synthetic spectrum from the
model atmospheres used in Figure 9b.  The strong allowed, 
semi-forbidden, and forbidden lines 
in the spectrum arise from the optically thin ejecta
beyond the largest radii considered in the model atmospheres.
The pseudo-continuum is well reproduced by both synthetic 
spectra but the LTE spectrum shows very strong 
Fe\,{\sc ii} emission lines at 2410 {\AA}, 2640 {\AA}, 
and 2780 {\AA} (Figure 10a).  The Fe\,{\sc ii} emission in the 
wing of the Mg\,{\sc ii} 2800 {\AA} produces a very strong line 
which equals the intensity of Mg\,{\sc ii} observed in OS And 86.  
The NLTE synthetic spectrum shows 
none of these Fe\,{\sc ii} emission features and a weak 
Mg\,{\sc ii} line in Figure 10b.  The Mg\,{\sc ii} line
is further reduced in strength if the model atmosphere is reduced 
in magnesium abundance by a factor of ten.  
Additional Mg\,{\sc ii} emission is produced from the optically 
thin ejecta beyond the outer model radius.  
In order to produce the features in the pseudo continuum seen in the
IUE spectrum, Fe\,{\sc ii} {\it must} be treated in NLTE.  The LTE
models overpopulate Fe\,{\sc ii} resulting in very strong 
Fe\,{\sc ii} lines that are not observed in the IUE spectrum.

\section{Summary}

OS And 86 was a fast CO type nova whose ejecta were optically thick in
the UV for about one month after visual maximum.  This fact, along with
the early {\it IUE} coverage, makes it an ideal candidate to determine
the chemical composition and physical conditions in the early nova outburst
by using a model atmosphere analysis.  In order to accurately model the
data, we require the reddening.  The 4 different methods used in this 
study give E(B-V) = 0.25 $\pm$ 0.05 as the reddening to OS And 86.
To determine the absolute properties, such as the luminosity, we 
have derived a distance to OS And 86 of 5.1 $\pm$ 1.5 kpc.  At this
distance and at a galactic latitude of -12.$^\circ$1 OS And 86 is
$\ge$ 1 kpc below the galactic disk. 

To model OS And 86 we created three synthetic spectral libraries 
(varying only the model temperature) with different abundance sets.
As a starting point, the first library contains solar abundance 
synthetic spectra.  Because of OS And 86's location in the galaxy and 
to investigate the possibility that OS And 86 may be a member of the
metal poor halo population, the second library consists of
metal poor synthetic spectra. 
Since earlier studies of Nova V1974 Cyg 1992 and Nova V705 Cas 1993
(Hauschildt et al. 1994a, 1994b) reported enhancements of the 
metals C,N,O and Fe relative to hydrogen we produced a set of metal rich 
synthetic spectra for the last library. 

The {\it IUE} spectra are best fit by synthetic spectra with solar 
metallicities although we can not rule out the metal poor synthetic
spectra with the {\it IUE} data alone.  The fact that we do not find
evidence for a metal enhancement is puzzling since theory
predicts that CNO elements should be enhanced relative 
to hydrogen caused by the mixing of white dwarf core material 
with accreted material (Starrfield 1989).  Further, the 
synthetic spectra are in better agreement at 2800 {\AA} when 
the Mg abundance is decreased from solar by an order of magnitude 
in all models.  
A possible explanation is that the secondary star of OS And 86 is
a galactic halo subdwarf.  A subdwarf would provide metal
poor material that was accreted onto the white dwarf.  Mixing
with the white dwarf core would enhance the CNO elements
but leave the high Z metals, e.g. Mg, Fe, etc., essentially unchanged.
Unfortunately, we were not able to
determine the CNO elemental abundances very accurately, but the 
``pre-nebular'' spectra show strong line emission from the CNO elements
indicating that these elements are overabundant with respect to 
solar.  We have shown that treating other elements, 
namely Fe\,{\sc ii}, in NLTE 
significantly improves the fits to the {\it IUE} spectra.  Future
work will include a more accurate determination of the CNO 
abundances using model atmospheres with CNO in NLTE (Hauschildt et al.
1996)

The model atmospheres give insight into the physical conditions 
of the outburst.  During the ``continuous mass loss'' phase, the models
show that the atmosphere is relatively ``cool'' and dense while the 
Fe\,{\sc ii} abundance is high.  The synthetic spectra show, and the
{\it IUE} spectra confirm, that the spectra at this epoch are dominated
by Fe\,{\sc ii} absorption.  As the nova shell expands, the electron
temperatures rise and the electron densities drop at a fixed 
location in space.  The density and opacity drop due to expansion
and the ejecta outside the model atmosphere are now exposed to a hotter
radiation field.  This leads to the
formation of the ``pre-nebular'' lines seen super-imposed on the hot
pseudo-continuum in the {\it IUE} spectra.

The model atmospheres also provide the bolometric flux.  
We use synthetic spectra to determine the flux of OS And 86 
at all wavelengths and our derived distance to obtain a bolometric 
luminosity of 5 $\pm 1 \times$ 10$^4$ L$_{\odot}$ or   
roughly the Eddington limit for a 1 solar mass white dwarf.  OS And
maintained a constant bolometric luminosity for 10 days after maximum
light in V.  After that time, the contribution from emission lines 
formed outside of the model atmosphere becomes significant and 
the bolometric luminosity calculated from our model
atmospheres drops.

\section{Acknowledgments}

It is a pleasure to thank J. Krautter, S. Pistinner, G. Shaviv, 
J. Truran, and R. Wade for stimulating discussions.  This work was 
supported in part by NASA and NSF grants to Arizona State University,
University of Oklahoma, and Wichita State University. 
Some of the calculations presented in this paper were performed on the
IBM SP2 of the Cornell Theory Center (CTC), and on the Cray C90 
of the San Diego Supercomputer Center (SDSC), supported by the NSF,
we thank them for a generous allocation of computer time.
The ultraviolet data were obtained with the International 
Ultraviolet Explorer Telescope and we gratefully acknowledge 
the support of the IUE observatory in obtaining these data.

\end{document}

%% file: mac.tex
  at 12.0 true pt %scaled \magstep1

\def\etal{{et al}}

\def\b{\beta}

\def\rout{\ifmmode{r_{\rm out}}\else\hbox{$r_{\rm out}$}\fi}
\def\tmax{\ifmmode{\tau_{\rm max}}\else\hbox{$\tau_{\rm max}$}\fi}
\def\tstd{\ifmmode{\tau_{\rm std}}\else\hbox{$\tau_{\rm std}$}\fi}
\def\vmax{\ifmmode{v_{\rm max}}\else\hbox{$v_{\rm max}$}\fi}
\def\muE{\ifmmode{\mu_{\rm E}}\else\hbox{$\mu_{\rm E}$}\fi} 
\def\pE{\ifmmode{p_{\rm E}}\else\hbox{$p_{\rm E}$}\fi} 
\def\bmax{\ifmmode{\b_{\rm max}}\else\hbox{$\b_{\rm max}$}\fi}
\def\kms{\hbox{$\,$km$\,$s$^{-1}$}}

\def\ang{\hbox{\AA}}

\def\Lsun{\hbox{$\,$L$_\odot$} }

\def\rout{\hbox{$r_{\rm out}$} }

\def\lstar{\ifmmode{\Lambda^*}\else\hbox{$\Lambda^*$}\fi} 
\def\Rop{\ifmmode{[R_{ij}]}\else\hbox{$[R_{ij}]$}\fi}

\def\Rji{\ifmmode{[R_{ji}]}\else\hbox{$[R_{ji}]$}\fi}
\def\Rstar{\ifmmode{[R_{ij}^*]}\else\hbox{$[R_{ij}^*]$}\fi}

\def\Rjistar{\ifmmode{[R_{ji}^*]}\else\hbox{$[R_{ji}^*]$}\fi}
\def\DRji{\ifmmode{[\Delta R_{ji}]}\else\hbox{$[\Delta R_{ji}]$}\fi}
\def\DRij{\ifmmode{[\Delta R_{ij}]}\else\hbox{$[\Delta R_{ij}]$}\fi}

\def\ns{\ifmmode{N_{\rm s}}          % Anzahl der tau-punkte
        \else\hbox{$N_{\rm s}$}\fi}

\def\mat#1{{\bf #1}}     % Macro fr Matrizen
\def\vek#1{{#1}}         % Macro fr Vektoren

\newcount\eqcount
\def
  \nummer{
    \global\advance\eqcount by 1
    (\the\eqcount)
  }

\def
  \numadv{
    \global\advance\eqcount by 1
  }

\def
   \numout#1{
     (\the\eqcount #1)
  }

\def\ivek#1#2{\ifmmode{\vek{I}^{#1}_{#2}}
        \else\hbox{$\vek{I}^{#1}_{#2}$}\fi}

\def\tmat#1#2{\ifmmode{\mat{t}^{#1}_{#2}}
        \else\hbox{$\mat{t}^{#1}_{#2}$}\fi}
\def\rmat#1#2{\ifmmode{\mat{r}^{#1}_{#2}}
        \else\hbox{$\mat{r}^{#1}_{#2}$}\fi}
\def\bvek#1#2{\ifmmode{\beta^{#1}_{#2}}
        \else\hbox{$\beta^{#1}_{#2}$}\fi}

\def\lp{\ifmmode{\lambda^+_\tau}           % lambda +
        \else\hbox{$\lambda^+_\tau$}\fi}
\def\lm{\ifmmode\lambda^-_\tau             % lambda -
        \else\hbox{$\lambda^-_\tau$}\fi}